# RANCANGAN INFRASTRUKTUR E-BISNIS BUSINESS INTELLIGENCE PADA PERGURUAN TINGGI

Spits Warnars H.L.H
Fakultas Teknologi Informasi, Universitas Budi Luhur
Jl. Petukangan Utara, Kebayoran Lama, Jakarta Selatan 12260, Indonesia
Telp (021) 5853753, Fax (021): 7371164
Email : spits@bl.ac.id, www.spits.8k.com

**Abstrak**

Dalam rangka persaingan dengan pesaingnya perguruan tinggi harus memperlengkapi infraksturnya dengan dukungan informasi teknologi. Manajemen tingkat sebagai pembuat keputusan membutuhkan sesuatu yang dapat mendorong sistem untuk berkompetisi dengan perguruan tinggi lainnya, mereka membutuhkan pengetahuan informasi teknologi yang dapat mendukung mereka untuk dapat memprediksi kedepan dan membantu keseluruhan sistem untuk meningkatkan pelayanan. Business Intelligence sebagai salah satu istilah sistem pengambilan keputusan yang dapat membantu manajemen dengan sesuatu yang bersifat dapat diprediksi dan diputuskan. Perguruan tinggi membutuhkan desain infrastruktur untuk membuat dasar yang kokoh untuk implementasi business intelligence yang akan diimplementasikan pada internet atau yang disebut e-business.

**Kata kunci** : Business Intelligence, e-bisnis, infrastruktur, teknologi informasi, Data Warehouse

**Abstract**

In order to compete with others, high education need complete their infrastructure with Information technology support. High level management as a decision maker need something that can boost the system to compete with other high education, they need IT knowledge that can support them to view the future and can help the whole system to improve their services. Business Intelligence is one of term of Decision Support System which can help the management by something that they can forecast and decide. High Education need infrastructure design to make good foundation for business intelligent implementation which will be implemented on internet or e-business.

**Keywords** : Business Intelligence, e-bisnis, infrastucture, information technology, Data Warehouse

## 1. PENDAHULUAN

Business Intelligence adalah proses-proses, alat bantu dan teknologi untuk mengubah data menjadi informasi dan informasi menjadi pemahaman serta rencana untuk menggerakkan aktivitas bisnis yang efektif. Business Intelligence apapun namanya merupakan sebuah Decision Support System atau diistilahkan dengan nama lain seperti Datawarehouse atau knowledge management [10][11]. Akan tetapi apapun istilhanya system tersebut adalah sistem yang dibangun untuk membantu manusia dalam pengambilan keputusan.

Bisnis saat ini beralih ke sistem dimana yang tidak lagi dilakukan secara manual namun mengarah menggunakan bantuan teknologi informasi. Bisnis mulai merambah ke internet dan biasanya disebut dengan istilah e-commerce atau e-business [10].

## 2. ANALISA MODEL BISNIS

Dalam menentukan model bisnis bagi sistem aplikasi yang akan dibangun, perguruan tinggi menentukan kriteria bagi aplikasi yang akan dibangun tersebut sebagai berikut [5]:

a) Memberikan kemudahan, keamanan, kecepatan dan kenyamanan bagi user dalam membantu user dalam menentukan keputusan
b) Investasi yang ditanamkan tidak terlalu tinggi tetapi manfaat yang diperoleh dapat langsung dirasakan.
c) Menjadi sumber pendapatan baru

Selanjutnya dilakukan analisa terhadap pengembangan aplikasi dengan menggunakan 3 kriteria seleksi konsep bisnis sebagai berikut[10]:

1) Potensi Pasar
   Kriteria ini digunakan untuk memastikan apakah potensi pemakai aplikasi ini memang nyata ada. Kuncinya adalah secara pasti dapat diketahui siapa konsumennya atau pemakai aplikasi ini dan apa yang mereka inginkan.
2) Daya Saing
   Daya saing adalah kriteria yang digunakan untuk mengukur sejauh mana perguruan tinggi dengan dukungan sistem yang akan dikembangkan ini dapat memenangkan persaingan.
3) Nilai Ekonomis
   Apakah Aplikasi yang akan dikembangkan ini menguntungkan dan memiliki potensi bisnis adalah aspek-aspek yang diperhitungkan untuk mengetahui nilai ekonomis dari sistem ini.

## 3. POTENSI PASAR

Dasar pengembangan *e-business* adalah karena adanya potensi pasar dalam hal ini adalah :

a. Manajemen tingkat atas (Dekan, Ketua Program Studi)
b. Dosen
c. Mahasiswa
d. Orang tua/Wali

Kebutuhan untuk mendapatkan pelayanan prima berupa kenyamanan, keamanan dan kecepatan adalah kualifikasi yang dapat dipenuhi oleh sistem yang akan di bangun. Berikut ini akan dijabarkan secara mendetail hal-hal yang akan didapat dari aktor potensi pasar ini

a. Manajemen tingkat atas (Dekan, Ketua Program Studi)
   mendapat laporan-laporan yang mendukung sistem pengambilan keputusan manajemen.
b. Dosen
   mendapat laporan-laporan yang mendukung system pengambilan keputusan penilaian mahasiswa, bimbingan akademik mendapatkan laporan-laporan yang berhubungan dengan kegiatan belajar-mengajar.
c. Mahasiswa
   memperoleh informasi nilai akhir, mendapat laporan-laporan yang mendukung sistem pengambilan keputusan jumlah mata kuliah dan sks yang diambil
d. Orang tua/wali
   Menyajikan informasi umum bagi orang tua mahasiswa baru untuk menentukan jurusan yang dipilih untuk anaknya. Orang tua mengontrol perkembangan nilai dan absensi kuliah anaknya setiap semester.

## 4. DAYA SAING

Untuk melihat posisi dan persaingan dari *e-business* dilakukan dengan menggunakan analisa untuk industri Porter Five Forces Competitive Model dari Michael Porter pada gambar dibawah ini [9].

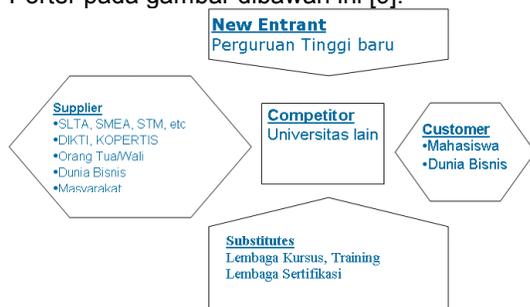

Gambar 1. 5 forces competition untuk perguruan tinggi

Berdasarkan analisa gambar diatas yang mana menggunakan five forces competitive Michael Porter maka dapat dipetakan ada 5 hal yang mempengaruhi persaingan yaitu :

**a. Competitor**

Merupakan pesaing dari dunia perguruan tinggi lain atau perguruan tinggi lain terutama yang menyelenggarakan pendidikan komputer. Pertumbuhan perguruan tinggi ini harus diperhatikan agar dapat diketahui siapa saja yang menjadi saingan dalam proses bisnis pendidikan ini. Perlu direncanakan strategi jangka panjang perusahaan untuk dapat bersaing dengan perguruan tinggi lain atau perguruan tinggi pesaing.

**b. Supplier**

Merupakan pihak-pihak yang bertindak sebagai penyedia baik mahasiswa sebagai subyek yang akan dididik maupun peraturan pemerintah yang mendukung kegiatan pendidikan di Indonesia. Pihak-pihak supplier tersebut diantaranya :

- Sekolah menengah atas seperti SMA, SMEA, STM dan lain-lain. Merupakan supplier terbesar yang menyediakan calon mahasiswa yang akan menjadi kustomer dunia pendidikan perguruan tinggi. Merupakan tuntutan dunia kerja saat ini untuk mempunyai latar belakang pendidikan yang memadai, pendidikan menengah atas belumlah cukup untuk memenuhi persaingan dunia kerja saat ini. Apalagi dengan semakin banyaknya institusi perguruan tinggi dan biaya yang semakin terjangkau. Pendidikan pada sekolah menengah atas ini mempengaruhi peluang yang timbul untuk mendapatkan mahasiswa baru. Kerja sama dengan pihak sekolah menengah atas ini dibutuhkan untuk selalu menarik lulusan sekolah menengah atas tersebut untuk menjadi kustomer.
- Peraturan pemerintah yang secara khusus untuk perguruan tinggi ditangani oleh DIKTI (Dirjen Pendidikan Tinggi) dan KOPERTIS (Koordinator Perguruan Tinggi). Pihak DIKTI dan KOPERTIS ini yang bertanggung jawab untuk mengelola dan mengatur perguruan tinggi agar dapat berjalan pada aturan-aturan pendidikan yang telah ditetapkan oleh pemerintah melalui Departemen Pendidikan Nasional. Pihak perguruan tinggi harus memperhatikan aturan-aturan yang telah ditetapkan agar dapat bersaing, bertahan dan menang dalam persaingan.
- Masyarakat, Orang tua/wali, Dunia bisnis. Pandangan / "*image*" masyarakat sangat dibutuhkan agar menjadi suatu "*trade mark*" dan dikenal masyarakat bahwa institusi pendidikan tersebut adalah benar-benar berkualitas dan memenuhi standar keinginan masyarakat. Dimana orang tua/wali yang merupakan bagian dari masyarakat tersebut merupakan sebagai pemicu untuk menentukan apakah anaknya diinginkan untuk dididik di suatu institusi perguruan tinggi. Selain itu dunia bisnis yang terkini perlu selalu diperhatikan oleh pihak perguruan tinggi untuk selalu menjaga mutu lulusan. Perguruan tinggi harus selalu memperhatikan kebutuhan dunia bisnis saat ini dalam dunia kerja dan kemungkinan dunia kerja yang ada. Dunia bisnis ini juga merupakan bagian dari masyarakat yang selalu mengontrol dan memberikan masukan pada perguruan tinggi.

**c. New Entrant**

Merupakan pendatang baru dalam dunia pendidikan tinggi. Pihak perguruan tinggi harus selalu memperhatikan dan mewaspadai pertumbuhan perguruan tinggi baru, apalagi dengan diperbolehkan pihak perguruan tinggi di luar Indonesia untuk mengembangkan sayapnya di negara ini merupakan perhatian mendalam dari pihak perguruan tinggi untuk selalu menjaga mutu didikan.
**d. Customer**
Merupakan individu yang menjadi sumber pendapatan dan target sasaran bisnis pendidikan tinggi. Pihak perguruan tinggi harus selalu memperhatikan dan memuaskan mahasiswa sebagai pemakai jasa pendidikan tinggi ini. Kepuasan yang didapat oleh mahasiswa merupakan sebuah investasi yang mahal bagi pertumbuhan perguruan tinggi. Mahasiswa-mahasiswa yang puas tersebut akan menjadi iklan yang berjalan dan hidup dan secara terus-menerus akan menjadi indikator kenaikan jumlah mahasiswa.
**e. Subtitutes**
Merupakan pengganti yang akan mempengaruhi mengurangnya jumlah mahasiswa yang menjadi sumber pemasukan bagi perguruan tinggi. Institusi pengganti tersebut adalah seperti lembaga kursus, training dan pelatihan serta lembaga sertifikasi. Bila tidak diperhatikan secara seksama lembaga pengganti ini akan bertumbuh menjadi *"new entrant"*. Selain itu pihak perguruan tinggi perlu memikirkan bagaimana agar perguruan tinggi dapat masuk juga kedalam bidang pengganti ini.

## 5. NILAI EKONOMIS
Dengan adanya business intelligence pada perguruan tinggi keuntungan ekonomis yang dapat diperoleh antara lain:
1) Menciptakan lulusan yang berkwalitas dan tepat waktu dengan nilai IPK yang memuaskan.
2) Menciptakan lingkungan kerja yang kondusif dengan efisiensi biaya dan waktu, misalkan mengurangi pertemuan face-toface antara orang tua/wali dengan pihak perguruan tinggi (dosen, manajemen dan karyawan).
3) Mengurangi jumlah karyawan, terutama yang berkaitan dengan pembuatan laporan-laporan pengambilan keputusan. (efisiensi biaya).
4) Mengurangi biaya kertas dan cetak (tinta printer) dan biaya penggandaan (tinta foto copy).
5) Menarik minat siswa/i sekolah menjadi mahasiswa.
6) Waktu pengambilan keputusan menjadi lebih cepat.
7) Waktu pembuatan laporan atau menghasilkan informasi menjadi lebih cepat.
8) *Image* atau *brand* Perguruan tinggi akan menjadi lebih baik.

Selain keuntungan ekonomis yang didapat dengan adanya business intelligence pada perguruan tinggi akan ada pula nilai tambah yang ditawarkan diantaranya :
1) Bagi Manajemen tingkat atas (Dekan, Ketua Program Studi);
    a. Mendapatkan tampilan laporan yang mudah dimengerti.
    b. Memperoleh laporan secara cepat dengan langsung mengakses, (ada aplikasinya)
2) Bagi Dosen
    a. Mendapatkan tampilan laporan yang mudah dimengerti.
    b. Membantu Dosen dalam membimbing kegiatan akademik mahasiswa bimbingannya
3) Bagi Mahasiswa
    a. Mendapatkan tampilan laporan yang mudah dimengerti
    b. Membantu mahasiswa lulus tepat waktu dengan nilai IPK memuaskan
4) Bagi Orang tua/wali
    a. Mendapatkan tampilan laporan yang mudah dimengerti
    b. Memudahkan pengontrolan perkembangan pendidikan anaknya.

Penurunan rancangan e-bisnis *business intelligence* dari model bisnis ke arsitektur dijembatani oleh rancangan proses-proses bisnis. Arsitektur adalah rancangan infrastruktur untuk menjalankan proses-proses bisnis. Adapun arsitektur tersebut terdiri dari [6]:
1) Arsitektur Konseptual
2) Arsitektur Logis
3) Arsitektur Eksekusi/Fisik

## 6. ARSITEKTUR KONSEPTUAL
Merupakan struktur dan interaksi antar aktor yang terlibat dalam proses bisnis yang digambarkan dengan menggunakan use case diagram sebagai salah satu diagram Unified Modeling Language. Proses-proses bisnis ini adalah sarana untuk merealisasikan strategi bisnis. Selain itu memperlihatkan teknologi-teknologi kunci yang akan digunakan.

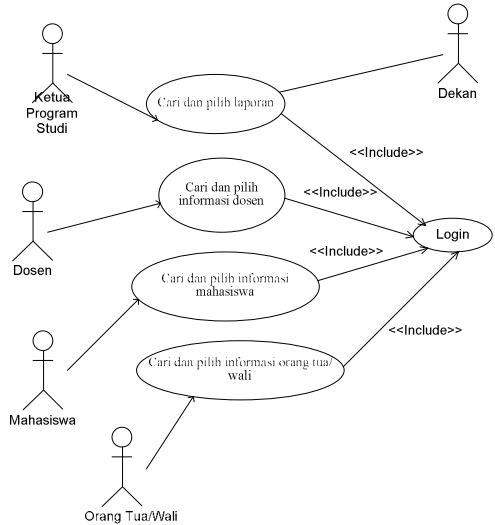

Gambar 2. Use Case e-bisnis business Intelligence perguruan tinggi

**Use case specification**
Use Case : Cari dan pilih Laporan
Aktor yang terlibat: Dekan, Ketua Program Studi
Keterangan : <<include>>
Use case : Login
Keterangan : Aktor memasukkan kode identitas dan password yang telah diberikan. Sistem akan mencari pada database anggota dan akan memberikan validasi. Apabila kode identitas dan password sesuai maka aktor dapat melanjutkan ke proses selanjutnya, jika tidak maka akan ditolak.
 Apabila validasi login terpenuhi maka Dekan dan Ketua Program Studi dapat menentukan dan memasukkan pilihan-pilihan data yang akan didapatkan yang telah disesuaikan dengan rancangan layar untuk proses mencari dan pilih laporan. Sesuai dengan pilihan-pilihan data yang diinput sistem akan menampilkan pilihan laporan/informasi dalam bentuk grafik atau laporan dan dapat dicetak pada printer.

Use case : Cari dan pilih informasi dosen
Aktor yang terlibat : Dosen
Keterangan : <<include>>
Use case : Login
Keterangan: Aktor memasukkan kode identitas dan password yang telah diberikan. Sistem akan mencari pada database anggota dan akan memberikan validasi. Apabila kode identitas dan password sesuai maka aktor dapat melanjutkan ke proses selanjutnya, jika tidak maka akan ditolak.
 Apabila validasi login terpenuhi maka Dosen dapat menentukan dan memasukkan pilihan-pilihan data yang akan didapatkan yang telah disesuaikan dengan rancangan layar untuk proses mencari dan pilih informasi dosen. Sesuai dengan pilihan-pilihan data yang diinput sistem akan menampilkan pilihan laporan/informasi dalam bentuk grafik atau laporan dan dapat dicetak pada printer.

Use case : Cari dan pilih informasi mahasiswa
Aktor yang terlibat : Dosen
Keterangan : <<include>>
Use case : Login
Keterangan: Aktor memasukkan kode identitas dan password yang telah diberikan. Sistem akan mencari pada database anggota dan akan memberikan validasi. Apabila kode identitas dan password sesuai maka aktor dapat melanjutkan ke proses selanjutnya, jika tidak maka akan ditolak.

Apabila validasi login terpenuhi maka Dekan mahasiswa dapat menentukan dan memasukkan pilihan-pilihan data yang akan didapatkan yang telah disesuaikan dengan rancangan layar untuk proses mencari dan pilih informasi mahasiswa. Sesuai dengan pilihan-pilihan data yang diinput sistem akan menampilkan pilihan laporan/informasi dalam bentuk grafik atau laporan dan dapat dicetak pada printer.

Use case : Cari dan pilih informasi orang tua/wali
Aktor yang terlibat: Dosen
Keterangan : <<include>>
Use case      : Login
Keterangan : Aktor memasukkan kode identitas dan password yang telah diberikan. Sistem akan mencari pada database anggota dan akan memberikan validasi. Apabila kode identitas dan password sesuai maka aktor dapat melanjutkan ke proses selanjutnya, jika tidak maka akan ditolak.

Apabila validasi login terpenuhi maka Orang tua/Wali dapat menentukan dan memasukkan pilihan-pilihan data yang akan didapatkan yang telah disesuaikan dengan rancangan layar untuk proses mencari dan pilih informasi Orang Tua/Wali. Sesuai dengan pilihan-pilihan data yang diinput sistem akan menampilkan pilihan laporan/informasi dalam bentuk grafik atau laporan dan dapat dicetak pada printer.

Berdasarkan use case diagram diatas maka diidentifikasi event-event dalam setiap proses bisnis dan respon yang diharapkan dari sistem pada use case list. Use case list ini memudahkan untuk menangkap berbagai situasi atau kasus yang mungkin terjadi dalam proses bisnis serta mengidentifikasi fungsi-fungsi komponen aplikasi yang memberikan respon terhadap event.

Ada 3 jenis tipe event bisnis yaitu :
1. Event eksternal, dibangkitkan oleh aktor eksternal dan berupa input data.
2. Event Temporal, dibangkitkan oleh suatu jadwal atau periode.
3. Event Perubahan Status, dibangkitkan oleh perubahan status atau kondisi internal sistem.

| No | aktor | event | trigger | response |
|---|---|---|---|---|
| 1. | Manajemen tingkat atas (Dekan dan Ketua Program Studi) | Cari dan pilih laporan | Input: data yang dibutuhkan (program studi dan alamat) | Tampilan daftar mahasiswa program studi lengkap dengan alamatnya. |
| | | | Input: program studi dan asal sekolah | Tampilan daftar mahasiswa menurut asal sekolah |
| | | | Input: program studi dan lama studi | Tampilan daftar mahasiswa berdasarkan pembagian lama studi (< 4 tahun, 4 – 5 tahun, > 5 tahun). |
| 2. | Dosen | Cari dan pilih informasi dosen | Input: IPK dan IPS | Tampilan daftar mahasiswa menurut IPS dan IPK-nya yang dibawah bimbingannya |
| | | | Input: Id_Dosen dan semester+tahun ajaran | Tampilan jadwal mengajar menurut semester dan tahun ajarannya |
| 3 | Mahasiswa | Cari dan pilih informasi mahasiswa | Input: NIM | Tampilan berupa nilai akhir, jumlah sks yang telah diambil dan prediksi kelulusan |
| | | | Input: Jurusan di SMA | Tampilan berupa laporan (IPK dan lama studi) tentang kelulusan alumni sesuai dengan jurusan di SMA-Nya |
| 4 | Orang tua/wali | Cari dan pilih informasi orang tua/wali | Input: Nama, program studi dan tahun masuk | Tampilan berupa Nama, NIM, program studi, IPK, nilai IP, mata kuliah dan absensi per semester. |
| | | | Input: Jurusan di SMA | Tampilan berupa laporan (IPK dan lama studi) tentang kelulusan alumni sesuai dengan jurusan di SMA-Nya |

Tabel 1. Use Case List

## 7. ARSITEKTUR LOGIS

Menggambarkan proses bisnis dan aliran data dan spesifikasi interface. Selain itu rancangan arsitektur logis ini merupakan jembatan penghubung antara rancangan model bisnis dengan rancangan arsitektur fisik yang diidentifikasi dari faktor-faktor keberhasilan utama (Critical Success Factor). Rancangan arsitektur logis ini mengacu pada use case diagram dan dimodelkan dengan diagram arus data (Data Flow Diagram).

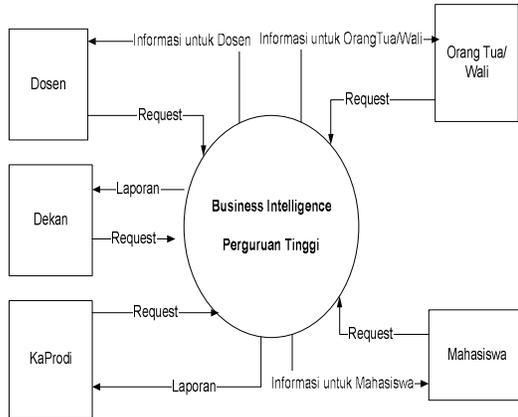

Gambar 3. Diagram Context e-bisnis business Intelligence perguruan tinggi

Berdasarkan use case list dibuatlah event diagram yang akan menjelaskan proses bisnis masing-masing event pada use case list.

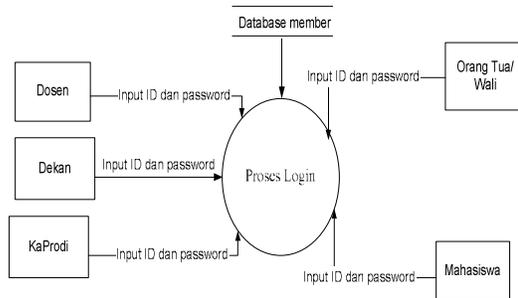

Gambar 4. Event Diagram proses login

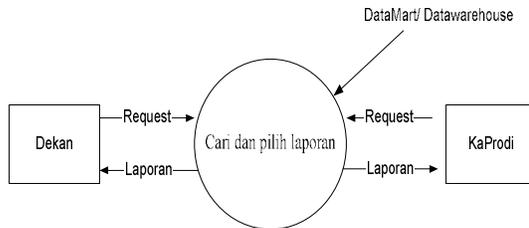

Gambar 5. Event Diagram proses Cari dan pilih Laporan

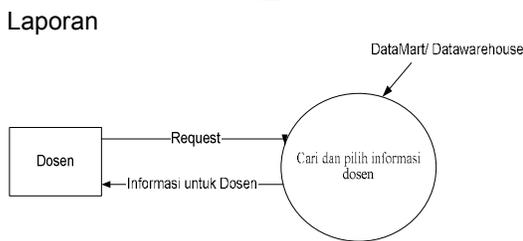

Gambar 6. Event Diagram proses Cari dan pilih Informasi Dosen

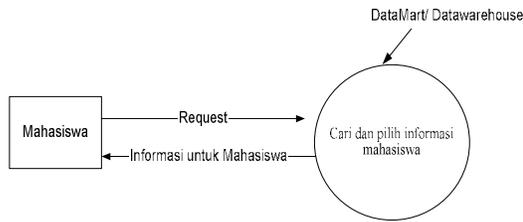

Gambar 7. Event Diagram proses Cari dan pilih Informasi Mahasiswa

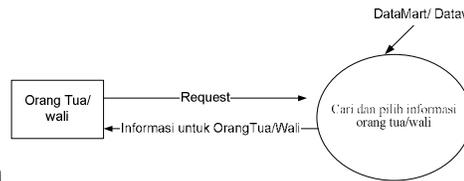

Gambar 8. Event Diagram proses Cari dan pilih Informasi Orang Tua/Wali

Berdasarkan event-event diagram diatas dibentuklah diagram aliran data sistem yang merupakan rancangan arsitektur logis. Diagram aliran data sistem ini menggambarkan secara keseluruhan database yang digunakan, aktor-aktor yang terlibat atau ekternal entitas yang terlibat dan berikut komponen-komponen aplikasi sistem. Diagram aliran data sistem ini seringkali juga disebut sebagai diagram overview.

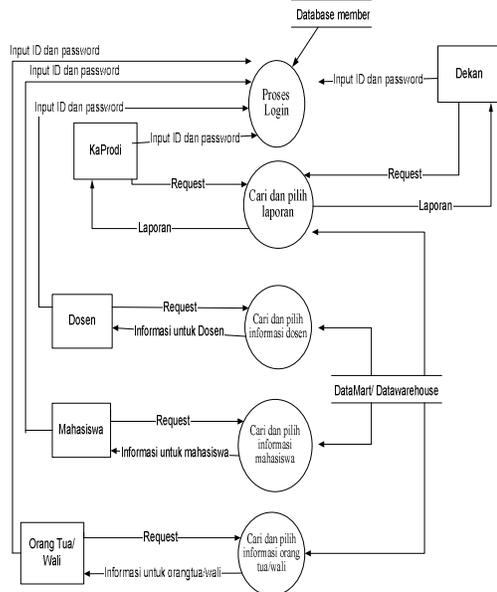

Gambar 9. Rancangan arsitektur logis e-bisnis business Intelligence perguruan tinggi

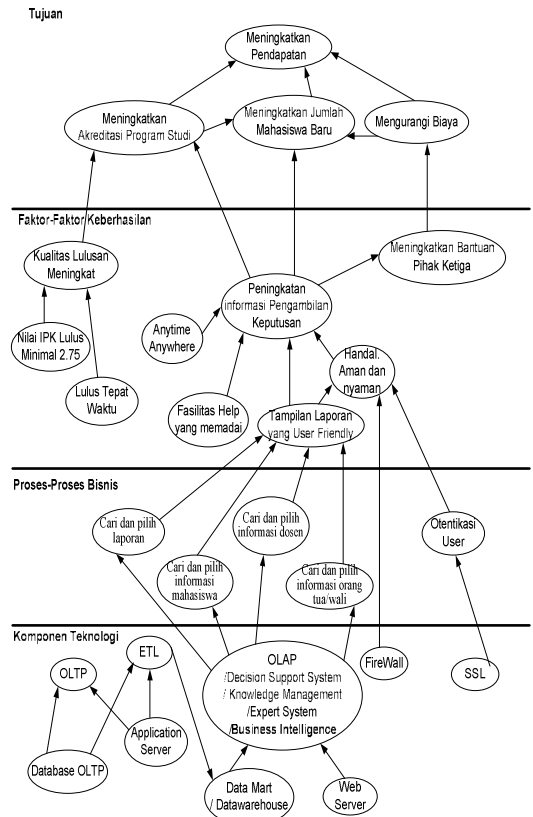

Gambar 10. Faktor Keberhasilan Utama (CSF) e-bisnis business Intelligence perguruan tinggi

## 8. ARSITEKTUR EKSEKUSI/FISIK

Menggambarkan topologi/struktur komponen-komponen implementasi. Rancangan arsitektur aplikasi ini dipetakan ke rancangan infrastruktur dengan memperhatikan aspek jaminan tingkat layanan (Services Level Agreement) dan aspek pengembangan lanjut. Yang terdiri dari :
1) Reliability
2) Performance
3) Scalability dan Extentibility

1) Reliability
   Reliability yang diharapkan dari sistem yang akan dikembangkan ditinjau dari beberapa hal atau aspek, yaitu:
   a. Ketersediaan (Availability)
      Availability sistem yang dibangun adalah 100% yaitu 24 jam x 7 hari atau non stop karena sistem bersifat online dengan menggunakan teknologi Internet.
   b. Akurasi (Accuracy)
      Akurasi dari sistem, khususnya terkait dengan fungsionalitas penyediaan informasi kepada pengguna dan pencarian data harus 100%.
   c. Mean Time Between Failure (MBTF)
      Mean Time Between Failure yang diharapkan dari sistem adalah 6 bulan sehingga sistem jarang mengalami kegagalan.
   d. Mean Time To Repair (MTTR)
      Mean Time To Repair yang diharapkan dari sistem adalah 1x24 jam sehingga jika ada kegagalan maka toleransi waktu untuk memperbaikinya adalah 1x24 jam sehingga hari berikutnya sistem sudah tersedia dan dapat digunakan kembali.
2) Performance

Performance yang diharapkan dari sistem yang akan dikembangkan ditinjau dari beberapa hal atau aspek, yaitu:
a. Response Time
Response time yang diharapkan dari sistem yang dibangun adalah maksimal 10 detik untuk proses loading halaman web dan pengelolaan atau penyediaan informasi maupun content.
b. Throughput
Throughput yang diharapkan dari sistem yang dibangun adalah mampu melayani minimal 10 transaksi per detik.
c. Capacity
Capacity yang diharapkan dari sistem yang dibangun adalah mampu melayani minimal 100 pengguna pada saat yang bersamaan.
d. Flexibility
Flexibility yang diharapkan dari sistem yang dibangun adalah web browser independent, artinya sistem dapat diakses dengan menggunakan web browser yang berbeda-beda (Internet Explorer, Netscape, dll)..

3) Scalability dan Extentibility
Scalability yang diharapkan dari sistem yang akan dikembangkan adalah sistem mampu menangani jumlah pengguna yang selalu bertambah. Selain itu, sistem juga diharapkan mampu untuk mengantisipasi jika akan ada penambahan layanan maupun teknologi baru (extentibility).

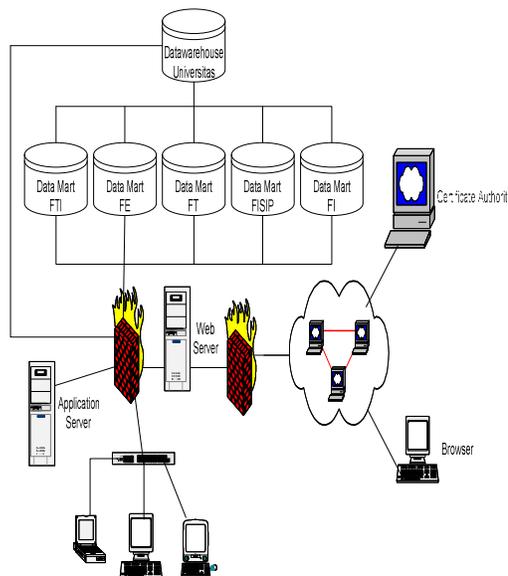

Gambar 11. Racangan arsitektur Eksekusi/fisik e-bisnis business Intelligence perguruan tinggi

**KESIMPULAN**

Analisa dan desain untuk masing-masing implementasi business intelligence ini akan sangat tergantung pada team yang ditunjuk oleh perguruan tinggi. Antar satu perguruan tinggi dengan perguruan tinggi lainnya mempunyai keunikan masing-masing yang bisa saja berbeda dalam proses bisnis dan implementasi penerapan teknologi informasi. Dukungan manajemen tingkat atas jelas sekali sangat dibutuhkan, selama manajemen tingkat atas tidak memandang teknologi informasi sebagai sebuah alat yang dapat membantu dalam persaingan dengan perguruan tinggi lain maka akan sia-sia penerapan business intelligence ini. Dukungan dari semua manajemen sangat dibutuhkan untuk kelancaran implementasi dan pengembangan kedepan.

Pada prinsipnya perguruan tinggi yang dibahas disini adalah perguruan tinggi swasta, namun tidak tertutup kemungkinan perguruan tinggi negeri dapat menerapkan dan mengimplementasikan business intelligence ini. Perguruan tinggi negeri cenderung "diatas angin" karena selalu menjadi favorit, dimana eksistensi yang sudah lama, dosen yang berpengalaman dan menjadi incaran calon mahasiswa oleh karena banyaknya saingan yang mendaftar dan kuliah dengan biaya yang terjangkau.

**About writer :**
Spits Warnars, H.L.H, was born in Semarang on April 21, 1972. Graduation for Computer Bachelor (S.Kom) degree in 1995 at Information System Study Program, Faculty of Information Technology, University of Budi Luhur (www.bl.ac.id). In Jan 2007 was graduated from Faculty of Computer Science, University of Indonesia (www.ui.ac.id) with Magister of Information Technology title (M.TI). Since September 2008 as P.hd Computer Science at Manchester Metropolitan University (MMU.ac.uk) under supervision Prof. Edmond Prakash(E.prakash@mmu.ac.uk).
Have been a lecturer at faculty of information technology, university of Budi Luhur since 1996 and as untemporal lecturer at department of Computer Science, University of Bina Nusantara (www.binus.ac.id)  since Jan 2008. Have been a reviewer for World Scientific and and Engineering Academy and Society, WSEAS (www.wseas.org) since 2006.